# Electron pairing in the pseudogap state revealed by shot noise in copper-oxide junctions

Panpan Zhou[1], Liyang Chen[2], Yue Liu[3], Ilya Sochnikov[4], Anthony T. Bollinger[5], Myung-Geun Han[5], Yimei Zhu[5], Xi He[6], Ivan Božović[5,6,*] & Douglas Natelson[1,*]

[1]Department of Physics and Astronomy, Rice University, Houston TX 77005, USA

[2]Applied Physics Graduate Program, Rice University, Houston, TX 77005, USA

[3]School of Physics, Peking University 100871, Beijing, China

[4]Department of Physics, University of Connecticut, Storrs CT 06269, USA

[5]Brookhaven National Laboratory, Upton, NY 11973 USA

[6]Department of Applied Physics, Yale University, New Haven CT 06520 USA

**In the quest to understand high-temperature superconductivity in copper oxides, a vigorous debate has been focused on the pseudogap — a partial gap that opens over portions of the Fermi surface in the 'normal' state above the bulk critical temperature ($T_c$).[1] The pseudogap has been attributed to precursor superconductivity, to the existence of preformed pairs, or to competing orders such as charge-density waves.[1-4] A direct determination of the charge of carriers as a function of temperature and bias could help resolve among these alternatives. Here, we report measurements of the shot noise of tunneling current in high-quality $La_{2-x}Sr_xCuO_4/La_2CuO_4/La_{2-x}Sr_xCuO_4$ (LSCO/LCO/LSCO) heterostructures fabricated using atomic-layer-by-layer molecular beam epitaxy, for several doping levels. The data delineate three distinct regions in the bias voltage-temperature ($V$-$T$) space. Well outside the superconducting gap region, the shot noise agrees quantitatively with independent tunneling of charge-$e$ carriers. Deep within the gap, shot noise is greatly enhanced, reminiscent of multiple Andreev reflections.[5-7] Starting above $T_c$ and extending to biases much larger than the gap, there is a broad region in which the noise substantially exceeds the expectations of single-charge tunneling, indicating pairing of carriers. Pairs are detectable deep into the pseudogap region of temperature and bias.**

The pseudogap has been detected in copper oxides and studied by many experimental probes, most directly by angle-resolved photoemission spectroscopy (ARPES)[8-11] and tunneling.[12,13] However, its microscopic origin, and its relation to other anomalous normal state properties and to high-temperature superconductivity (HTS), have remained the subject of much speculation.

One candidate idea is that the pseudogap is a high-temperature precursor of the superconducting state. In this scenario, at $T_c$ the global phase coherence is destroyed by thermal fluctuations, while preformed pairs exist well above $T_c$ and up to some higher pairing temperature (which may not be sharply defined).[1,2] Indeed, ARPES,[8-11] tunneling,[12-14] and terahertz spectroscopy[15] data are consistent with superconducting fluctuations detectable up to 10-20 K above $T_c$. The range expands with the sensitivity of the probe; thus, Nernst effect[16] and torque magnetometry[17] detect the signatures of vortices and fluctuating diamagnetism up to even higher temperatures. Note that in all copper oxides, the superfluid density is very low; the phase stiffness temperature is roughly the same as $T_c$, even at optimal doping,[18] and hence, thermal phase fluctuations must be very large near $T_c$. Moreover, $T_c$ has been found to scale with the superfluid density and appears to be kinematically controlled,[18] in line with strong-coupling theories of HTS.[3,4] However, a direct and quantitative signature of hole pairing above $T_c$ has remained elusive.

Another popular scenario is a "two-gap" picture in which the pseudogap is distinct from the superconducting gap and originates from some other instability competing with superconductivity.[1,4] Candidates include charge-density waves, $d$-density waves, stripes, electronic nematicity (broken rotational symmetry in the electron fluid), etc.[1] Low-energy excitations out of such a state should be some collective modes, e.g., oscillations of the phase and amplitude of the order parameter (phasons and amplitudons).

Measuring the charge of mobile carriers in the pseudogap state could discriminate between these possibilities. A population of preformed pairs would manifest as an average effective charge $q^*$ larger in magnitude than the electron charge $e$, while lack of well-defined current-carrying quasiparticles would appear as a suppressed effective charge below $e$. The most direct experimental probes of charge are the measurements of Aharonov-Bohm oscillations in nano-rings, Coulomb blockade in nanoscale "dots", and shot noise in nanowires or tunnel junctions. The short inelastic mean-free-path of carriers in the copper oxides, in particular at temperatures above $T_c$, and the challenge of nanofabrication without damaging material properties, currently make the first three approaches extremely technically challenging. Measurement of shot noise in large-area planar tunnel junctions remains as the most feasible candidate to infer the charge of the carriers in bulk samples in the normal state.

Shot noise refers to the intrinsic current fluctuations that occur when discrete charge carriers are driven through a device. The intensity of shot noise $S_I$ is directly related to the charge of the carriers. Seminal experiments have employed shot noise to detect fractionally charged quasiparticles in the fractional quantum Hall effect,[19,20] electron pairing in superconductors, and multi-charge tunneling in higher order Andreev reflection processes.[6,7] Two very recent shot noise experiments on HTS copper oxides have also yielded important findings,[21,22] including direct evidence at high bias for local trapping of charge in the polarizable insulating layers that separate the conductive $CuO_2$ planes.[22]

We have performed shot noise measurements on LSCO-based tunnel junctions. Our key results are summarized in Figure 1, showing the inferred percentage of paired charges contributing to the tunneling current, $z$, as a function of the doping level $x$, temperature $T$ and bias $V$. For comparison, we also indicate (by red dash-dot lines) the superconducting gap region outside which one would expect $z = 0$ from the BCS theory for the measured values of $T_c$. Apparently, for all doping levels studied here, the contribution of pairs to the tunneling current extends well outside the superconducting gap scale and deep into the pseudogap regime. Precursor superconductivity above $T_c$ has been observed previously in photoemission, tunneling, and terahertz experiments.[8-13,15] Remarkably, in sharp contrast to low-$T_c$ superconductors,[6,7] we observe pair contributions to tunneling also into the pseudogap region, at energies well outside the superconducting gap region, both below and above $T_c$. This suggests that pairs are present at least in large portions of the parameter space dominated by the pseudogap. This is in line with the conjectured existence of a pair-density wave, an unusual condensed matter state anticipated in theory[23,24] and observed in recent STM experiments on copper oxides.[25] In what follows, we substantiate these observations and inferences.

Fabricating high-quality superconductor-insulator-superconductor (SIS) tunnel junctions with high-temperature superconductors is very challenging. Since the *c*-axis coherence length in cuprates is extremely short (just few Å), any attempt to observe coherence effects requires the interfaces in SIS trilayer structures to be perfect on an atomic scale. With state-of-the-art atomic-layer-by-layer molecular beam epitaxy (ALL-MBE) technique, cuprate SIS trilayers can be grown with atomically sharp interfaces and extremely narrow transitions in the leads.[26]

For the present study, we have used ALL-MBE to synthesize trilayer LSCO/LCO/LSCO films with LSCO doping level of $x$ = 0.10, 0.12, 0.14, and 0.15 (near-optimal doping). Transition temperatures in these films (Methods) are 28 K, 34 K, 37 K, and 38 K, respectively. Figure 2a shows a schematic cross section of the heterostructure. The LCO layer thickness is precisely controlled to be 2.0 nm (*i.e.*, 3 monolayers of LCO). Figure 2b shows a cross-section of an actual device imaged using a scanning transmission electron microscope (STEM). Energy dispersive x-ray spectroscopy and atomic-resolution electron-energy-loss spectroscopy were used for La, Sr, and Cu elemental mapping. The micrographs demonstrate remarkable crystalline perfection and atomically sharp interfaces, consistent with previous extensive STEM studies of cuprate films synthesized by ALL-MBE. Atomic-force microscopy also shows that the surfaces are atomically smooth, except for occasional steps due to substrate miscut (see Methods and Extended Data).

From these heterostructures, we have fabricated tunnel junction devices using photolithography. Figure 2d shows a schematic diagram of an example device. A portion of the insulating $Al_2O_3$ cover layer (cyan) is removed to show the epitaxial LSCO/LCO/LSCO heterostructure buried underneath. The LCO layer (red) is an antiferromagnetic Mott insulator and acts as the tunnel barrier.

Precision measurements of the bias dependence of the differential conductance, $G = dI/dV$, where $I$ is the current and $V$ is the voltage bias, were performed via standard lock-in techniques. Two tunnel junctions were measured at each LSCO doping level in the top and bottom superconducting electrodes. A normalized example is shown in Figure 2e for a device with nearly-optimal doping ($x$ = 0.15).

Our conductance data are qualitatively and quantitatively consistent with the literature. As in previous measurements on similar structures,[26] the zero-bias conductance of the junctions decreases with decreasing temperature. A strong non-Ohmic conductance suppression near zero bias emerges as $T$ is reduced through and below $T_c$ (see Extended Data Figure 2 for details), as expected for SIS tunneling. In Figure 2e, weak coherence peaks are resolved near $V = \pm 2\Delta/e$, where $\Delta$ is the inferred magnitude of the superconducting gap. In lower-doped samples, the coherence peaks are broad and not easily resolved, consistent with increased smearing of such features in incoherent SIS tunneling.[27] In line with prior results,[26] no supercurrent is observed in any of these devices down to $T$ = 20 mK. As $T \to 0$, instead of being exponentially suppressed as in *s*-wave SIS structures, the zero-bias conductance converges to about 20-30% of the normal state value.

From Figure 2e, the superconducting gap is $2\Delta_0 \approx 15$ meV, essentially the same as the value inferred earlier from Andreev reflections observed in point-contact tunneling.[27] However, the gap does not close at $T_c$ but stays nonzero and evolves through $T_c$ smoothly without any kinks. Next, unlike in conventional BCS superconductors, the *I-V* characteristic is not Ohmic even for bias $eV \gg 2\Delta_0$; instead, $G(V)$ keeps increasing with bias, and has an asymmetric V-shape (Extended Data Figure 2b). This is typical of tunneling into HTS copper oxides. As seen from Figure 2e, the superconducting gap is essentially electron-hole symmetric.

At each temperature we measure the noise spectra up to 300 kHz as a function of bias using a cross-correlation technique involving two independent low-noise amplifier chains.[20] The measured voltage fluctuations are transduced to current fluctuations via the device's differential resistance at each bias. Details are shown in Methods. Zero-bias noise agrees quantitatively with Johnson-Nyquist expectations based on the measured zero-bias conductance.

Within the single-electron Poissonian tunneling approximation, the noise power spectral density at finite temperature $T$ is expected to be $S_{I,e} = 2eI \coth(eV/2k_BT)$.[7] This reduces to the Johnson-Nyquist noise in the zero-bias limit, and accounts for the finite temperature smearing of the Fermi-Dirac distribution. This expression has been used in analyzing other SIS systems, including those exhibiting multiple Andreev reflections.[6]

Figures 3a-3d shows the measured noise intensity of an $x$ = 0.14 device with the red dashed line indicating the dependence expected for single-electron tunneling, $S_{I,e}$. At temperatures far above $T_c$ = 37 K, the measured noise value agrees with this expectation very well. As temperature approaches $T_c$ from above, the measured noise noticeably exceeds $S_{I,e}$. When the temperature falls below $T_c$, the excess noise above $S_{I,e}$ becomes increasingly pronounced. At the lowest temperatures in our system, the noise is nonmonotonic, with peak features at ±6 mV, approximately $\pm\Delta/e$, if the full width of the zero-bias suppression of conductance is interpreted as $4\Delta/e$. Corresponding differential conductance data is in Extended Data Figure 3.

We define the noise ratio as $S_I/S_{I,e}$, the ratio of measured noise to the single-electron tunneling expectation, and plot this in Figure 3e. At zero bias, the noise reduces to the Johnson-Nyquist level, and the noise ratio must approach 1, regardless of the charge of the carriers. At temperatures below $T_c$, the shot noise is enhanced greatly, with large noise ratios well above 1, see Figure 3e. The noise ratio is non-monotonic versus bias, increasing quickly with bias initially, reaching a maximum at the bias energy close to $\Delta$, and decreasing again at higher biases. These large noise ratios are qualitatively reminiscent of multiple Andreev reflections (MAR), in which noise is enhanced as charge tunnels through multielectron processes,[5-7] a resemblance discussed further in Methods and Extended Data. The noise enhancement is largest at low temperatures and decreases gradually as temperature approaches $T_c$. However, the noise ratio stays significantly above 1 even at temperatures well above $T_c$. Even more telling, both below and above $T_c$ the noise ratio remains larger than 1 up to biases larger by a factor of two or more than $2\Delta_0/e$.

The above findings, that the noise is enhanced even for $V > 2\Delta_0/e$ and/or $T > T_c$, are very robust; we have observed the same results in every device we have studied so far. However, the details vary and depend on the doping level, as illustrated in Extended Data Figure 6.

The most natural explanation of the enhanced noise is a paired-charge contribution to tunneling that starts already in the pseudogap phase, for $T$ well above $T_c$ and/or $V$ well above $2\Delta_0/e$, and evolves into higher-order processes below $T_c$ at biases within $2\Delta_0/e$. To quantify our results, from the measured $S_I$ we can extract the $T$- and $V$-dependent 'effective charge' $q^*$ defined via $S_I = 2q^*I \coth(q^*V/2k_BT)$. In a standard Bardeen-Cooper-Schrieffer (BCS) superconductor, $q^* = e$ outside the superconducting gap region enclosed by the $2\Delta(T)/e$ line that terminates at $T_c$, while at low bias and temperature, $q^* \approx 2e$ in the absence of higher-order processes and can be even larger if higher-order processes contribute to tunneling.[5-7]

In our samples, we observe $q^* > e$ well outside the $2\Delta(T)/e$ line. In that region, we make the phenomenological assumption to model a fraction $z$ of tunneling current $I$ as contributed by paired carriers. Within this model the shot noise is expected[6] to be $S_I = (1-z)2eI \coth(eV/2k_BT) + z\, 4eI \coth(eV/k_BT)$. The experimentally determined function $z(V,T)$ is shown in Figures 1a-1d for the doping levels $x = 0.10, 0.12, 0.14$, and $0.15$, respectively. Clearly, at every doping pairs are present far outside the superconducting gap region $2\Delta(T)/e$ that would be expected in a $d$-wave BCS superconductor with the corresponding value of $T_c$. We note that there is a difference between the fraction of $c$-axis tunneling current contributed by paired carriers and the fraction of all carriers that are paired. The actual pair density could be larger, since the tunneling probability for incoherent pairs may well be smaller than that for single electrons. Moreover, note that unlike in scanning tunneling microscopy, which is spatially localized, these atomically-flat, large-area tunneling structures favor conservation of the transverse ($a$-$b$ plane) quasi-momentum in the $c$-axis tunneling, which is dominated by carriers from the antinode portion of the Fermi surface,[28] where the pseudogap is maximal.[1]

Our tunneling conductance data delineate the superconducting-gap region, the boundary of which is consistent with previous observations of the phase-fluctuating superconductivity by THz spectroscopy.[15] This superconducting-gap region is clearly distinct from the pseudogap region identified outside of this boundary, suggesting that these are two different phases. On the other hand, the evolution of both the conductance and the enhanced noise between the normal state and the superconducting-gap regions is very smooth, without any kinks at the boundary. The key new finding here is that electron pairing, as detected through super-Poissonian shot noise, persists deep into the pseudogap state and at bias energy scales large compared to the apparent superconducting gap scale.

The presence of pairs above $T_c$ and in a bias regime expected to be dominated by the antinodal portion of the Brillouin zone constrains models of the pseudogap. While low superfluid density implies that thermal phase fluctuations must be strong, this alone can hardly account for pairing at energies large compared to the superconducting gap. A possibility to explore is a pair density wave.[23-25] It is also intriguing how this relates to electronic nematicity, the spontaneous breaking of the rotational symmetry in the electronic fluid detected in the pseudogap region in several copper oxides.[29,30]

**Acknowledgements**

Yangyang Zhang helped with TEM sample preparation and Shize Yang with STEM EDS data acquisition. We are also grateful to Patrick Lee, Steven Kivelson, Douglas Scalapino, Junichiro Kono, Matthew Foster, Tsz Chun Wu, Juan Carlos Cuevas, Peter Samuelsson, Adrian Gozar and Ilya Drozdov for illuminating comments and questions. The research at Brookhaven National Laboratory, including heterostructure synthesis and characterization and device fabrication, was supported by the U.S. Department of Energy, Basic Energy Sciences, Materials Sciences and Engineering Division. X.H. was supported by the Gordon and Betty Moore Foundation's EPiQS Initiative through Grant GBMF4410. The work at the University of Connecticut was supported by the State of Connecticut. The research at Rice University was supported by the US Department of Energy, Basic Energy Sciences, Experimental Condensed Matter Physics award DE-FG02-06ER46337. Some of the Rice noise measurement hardware acquired through NSF DMR-1704264.

**Author contributions**

XH and IB synthesized the HTS heterostructures. AB and PZ fabricated the tunneling and Hall bar devices. PZ, LC, and YL built the noise measurement system. PZ performed and analyzed the noise measurements. MGH and YZ conducted the TEM studies. DN and IB designed the experiments. IS performed differential conductance measurements down to dilution refrigerator temperatures. All authors contributed to writing the manuscript.

**Author information**

Reprints and permissions information is available at www.nature.com/reprints

**Competing financial interests**

The authors declare no competing financial interests.

**Correspondence** (*)

Correspondence and requests for materials should be addressed to natelson@rice.edu, bozovic@bnl.gov.

# Figure Legends

**Figure 1**. **The percentage of tunneling paired charges, $z$, as a function of doping level $x$, temperature $T$ and bias $V$, as inferred from shot-noise measurements on LSCO/LCO/LSCO tunnel junctions.** **a-d**, the data for doping levels $x = 0.10$, $0.12$, $0.14$, and $0.15$, respectively. Red dash-dot lines: the superconducting gap region outside which one would expect $z = 0$ from the BCS theory for the measured values of $T_c$. Green dashed line: $V = k_B T/e$. As $eV/k_B T \rightarrow 0$, discrimination of $z$ via noise measurements is not possible (see Methods and Extended Data). Grey region indicates where uncertainty in $z$ exceeds 0.5. For all doping levels, the contribution of pairs to the tunneling current extends well outside the superconducting region and into the pseudogap regime.

**Figure 2**. **LSCO/LCO/LSCO tunneling structures synthesized by ALL-MBE.** **a**, Film schematic: a tunneling barrier consisting of three molecular layers (1.5 unit cells) of undoped LCO is sandwiched between the bottom and the top superconducting LSCO electrodes. **b**, A high-resolution cross-section image of the actual device obtained by aberration corrected scanning transmission electron microscopy (STEM) and high-angle annular dark-field imaging (HAADF). **c,** Elemental maps of Sr (green) and La (red) obtained by atomic-resolution energy dispersive x-ray spectroscopy (EDS) and electron-energy-loss spectroscopy (HREELS), respectively, with overlaid white lines showing averaged line profiles. Yellow dashed lines indicate the boundaries of the undoped LCO layers. **d**, Device schematic: photolithography and etching are used to prepare vertical tunneling devices, 10 or 20 μm in diameter. **e**, Tunneling differential conductance data normalized to those at $T = 50$ K, $G_{norm} = (dI/dV)/(dI/dV)_{50K}$, as a function of the bias voltage, for a junction with nearly-optimally doped ($x = 0.15$) LSCO electrodes.

**Figure 3**. **Noise compared with single electron tunneling expectations**. **a-d**, For $x = 0.14$ doping, at high temperatures the measured noise (blue points with standard deviation error bars – see Methods) agrees well with that expected for single electron tunneling ($S_{I,e}$, red dashed line), with no adjustable parameters. As $T$ approaches $T_c$, noise is clearly in excess of $S_{I,e}$. When $T \ll T_c = 37$ K, noise is nonmonotonic with peaks at approximately the half-width of the zero-bias conductance suppression. **e**, The noise ratio $S_I/S_{I,e}$ at the same temperatures as in **a-d**. The excess noise above $S_{I,e}$ results in a noise ratio larger than 1. The thin blue line is a spline interpolation.

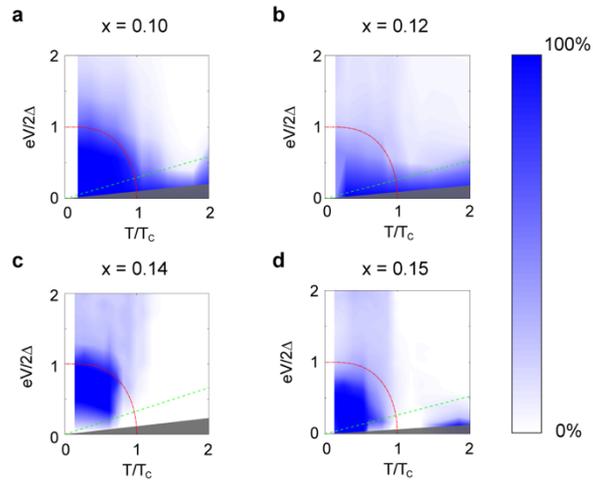

**Figure 1**

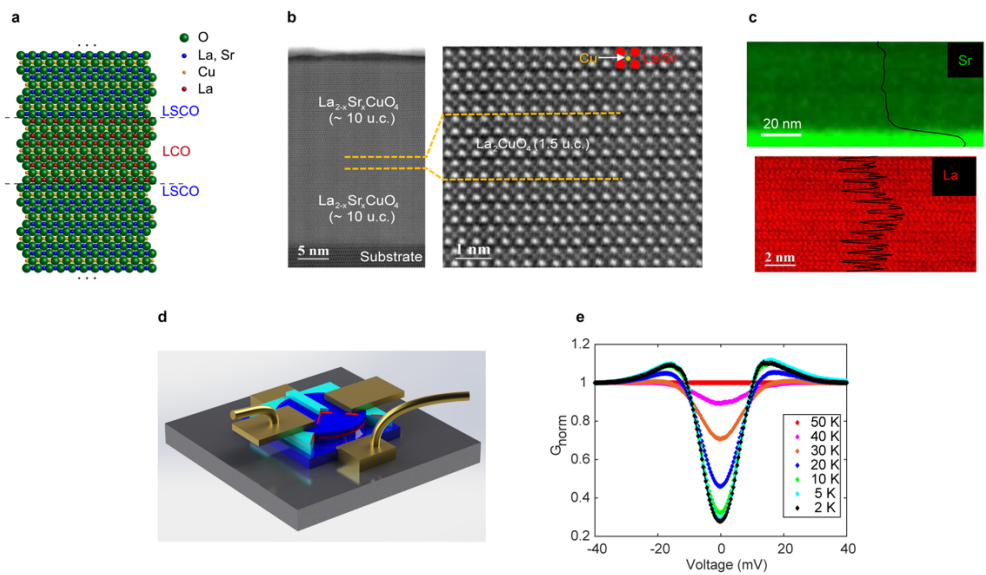

**Figure 2**

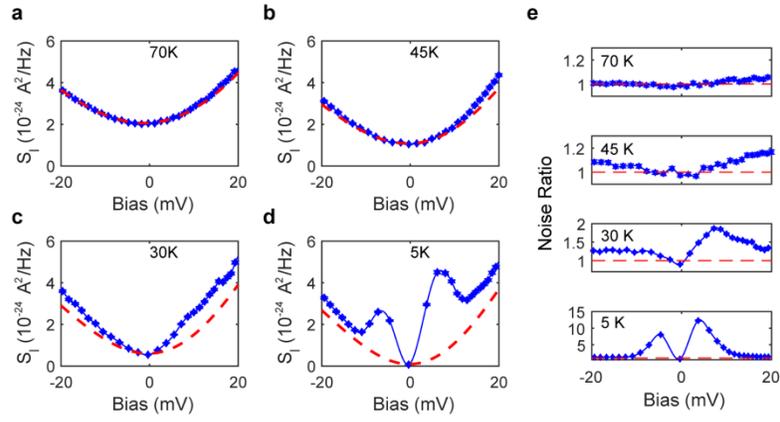

**Figure 3**

# Methods

## 1. Film growth and fabrication

The LSCO/LCO/LSCO heterostructures studied in this experiment were synthesized using an advanced atomic-layer-by-layer molecular beam epitaxy (ALL-MBE) system.[31-35] The film growth was monitored in real time by reflection high-energy electron diffraction (RHEED). The diffraction patterns provide information on the surface morphology and crystalline structure. The oscillations of the intensity of the specular reflection with time provide for a digital count of the number of deposited monolayers. Sufficient oxidation under high-vacuum conditions needed for MBE is accomplished using a source of pure ozone. The films were deposited on LaSrAlO$_4$ (LSAO) substrates polished perpendicular to the crystallographic (001) direction, so the epitaxy ensured that the CuO$_2$ planes in the LSCO films are parallel to the substrate surface on which they are grown. The substrate temperature was kept at about 650 ºC and the ozone partial pressure at about $2 \times 10^{-5}$ Torr.

The structure of the films was as follows. One monolayer of an overdoped LSCO was used as a buffer to nucleate the growth; then we deposited 39 monolayers of LSCO to serve as the bottom superconductor electrode, followed by the insulating barrier comprising 3 monolayers of LCO, and the top superconductor of 20 monolayers of LSCO. After the growth, a thin layer of Au (10 nm) was deposited on top of the film *in situ* to serve as a capping layer and protect the film surface. After the deposition, the films were annealed in vacuum for 30 minutes with the heater power reduced from 300 W to 50 W, corresponding to the sample temperature of about 250 ºC. We have established by experiments on many underdoped LSCO films that such annealing step is sufficient to remove interstitial oxygen from the LSCO electrodes as well as from the LCO barrier, rendering the later insulating[32-35] — as indeed verified by our *c*-axis transport measurements on fabricated devices.

The fabrication process for the tunnel junction devices is illustrated in Extended Data Fig. 1. Using the standard photolithography techniques, the LSCO/LCO/LSCO film was first milled with argon ions into small mesas (panel **b**). A second lithography step and ion milling etched away part of the top LSCO layer and the middle LCO layer. The etching depth is finely controlled to expose the bottom LSCO layer but not fully etch to the substrate (panel **c**). A thick layer of Al$_2$O$_3$ (100 nm) is evaporated to a photolithographically defined area to help isolate the top and bottom contacts (panel **d**). Finally, Au (150 nm) is evaporated to make the top and bottom contacts (panel **e**). Panel **f** shows a false-colored scan electron microscopy image of a fabricated device with a scale bar of 10 μm.

2. **Transmission electron microscopy characterization**

TEM samples were prepared by a Focused Ion Beam (FEI Helios Nanolab) using 2.0 keV $Ga^+$ ion for final milling. A focused 0.5 kV $Ar^+$ ion beam (Nanomill, Fischione Instruments, Inc.) was used to remove FIB damaged layers at liquid nitrogen temperature. For HAADF STEM images, a JEOL ARM 200CF equipped with a cold field emission source and two aberration-correctors at the Brookhaven National Laboratory was used with 200 keV electrons and the collection angles in the range of 67 to 275 mrad. For EELS spectrum imaging, La L edges (832 eV) were recorded with 0.1 eV/channel energy dispersion. The EELS acquisition time was 0.05 s/pixel with 0.039 nm per pixel. The convergent and collection semi-angles were 20 and 10.42 mrad, respectively. For Sr elemental mapping, a FEI Talos F200X equipped with a four-quadrant 0.9-sr energy dispersive X-ray spectrometer operated at 200 keV was used. Sr L (1.806 keV) signals were collected with acquisition time of ~ 3 mins with 0.6 nm pixel size. To enhance signal-to-noise ratio, principal component analysis was performed. Line profiles of La L edges were obtained after background and baseline subtractions. The 1.5 unit cell undoped LCO layer in the SIS architecture was not resolved in HAADF, EELS mapping using Sr L edge (1940 eV) and EDX mapping of the La edge due to low concentration (8 %) difference of Sr, but clearly visible in EDX mapping of Sr edge and EELS mapping of La edge.

3. **Film and device characterization**

On each chip, multiple Hall bar devices were also fabricated alongside the tunnel junctions. These Hall bar devices were used for measuring $T_c$ of both bottom and top LSCO layers at each doping. The measured $T_c$ temperatures for $x$ = 0.15, 0.14, 0.12 and 0.1 were 38 K, 37 K, 34 K and 28 K, respectively, which is in good agreement with previous reports on ALL-MBE grown LSCO film samples.[32-35] Mutual inductance measurements[18] on the as-grown multilayer films showed that the transition temperatures of the bottom and top LSCO layers were identical to within the width of the transition.

Precise measurements of the tunneling differential conductance were performed using standard lock-in amplifier techniques. A small ac voltage bias was superposed on top of a variable dc bias applied to the junction. The ac/dc voltage across the junction and ac/dc current driven through the junction were measured by a digitizer (NI-6521) and lock-in amplifiers (Signal Recovery 7265 and 7270). Extended Data Figure 2**b** shows a typical tunneling conductance of an $x = 0.15$ doped LSCO junction at temperatures below 50 K. As expected, the tunneling conductance is very nonlinear at low temperatures, especially below $T_c$, due to the gapping of quasiparticle excitations over much of the Fermi surface in the superconducting state. The systematic asymmetry in the differential conductance about zero bias was observed in all tunneling devices over a broad temperature range including above $T_c$. We ascribe this to a difference between the upper and lower LSCO layers in the epitaxial strain originating from the substrate, in combination with the polar nature of the material.[36]

For traditional *s*-wave superconductors, the BCS model predicts that at low temperatures, the zero-bias tunneling conductance will be suppressed exponentially to 0 as $T \to 0$. In the tunneling devices under study here, the zero-bias tunneling conductance converges to about 20-30% of the normal state value down to temperatures as low as 20 mK.

A key concern regarding tunneling structures is the uniformity of the tunneling barrier, and the possible presence of "pinholes" in such a thin barrier. Prior studies[26,37] of ALL-MBE grown LSCO/LCO/LSCO heterostructures have focused on this as a primary technical issue. There it was shown that even a 1 UC (1.3 nm) thick LCO barriers had no pinholes and were insulating. To be on the safe side, the devices for the present work had barriers 50% thicker (1.5 UC = 2 nm). These are likewise insulating, with no sign of pinholes that would short the junctions. In particular, the lack of any measurable supercurrent down to dilution refrigerator temperatures and picoamp resolution in all devices examined argues that there are no true pinholes (Extended Data Figure 2c). True pinholes would likely support supercurrent. Earlier work[37] has shown that severely underdoped LSCO can support proximity-induced superconductivity over distances much larger than the barrier thickness in these devices. Consistency of tunneling conductance from device to device also suggests limited variability in barrier transparency.

There are additional possible sources of variation in barrier properties. One is the substrate miscut from the ideal crystallographic plane orthogonal to the [001] direction, inevitable but varying from one substrate to another, which leaves surface steps that are transmitted to the substrate and generate antiphase dislocations; another is some Sr interdiffusion across the ideal geometric interface. This issue has already been examined in detail,[32,33,36] and it was found that Sr can diffuse within one LSCO layer (0.5 UC = 0.66 nm thick). In the present work, the only indication of possible variation in barrier transparency is the device-to-device variation in the magnitude of the maximum low-temperature noise enhancement.

## 4. Shot noise measurement circuit

The present experiments use a cross-correlation method to measure the shot noise. Extended Data Figure 4**a** shows a schematic electrical circuit diagram of the experimental setup. A tunable voltage source (NI-DAQ6521) is heavily filtered with *LC* filters (>60 dB attenuation at frequency beyond 100 Hz) to provide a clean bias. Two larger resistors (~200 kΩ each) are used to limit the input current. The sample is loaded inside a cryostat (PPMS from Quantum Design) using a home-built shot noise probe with careful shielding and isolation from the PPMS ground and environment. The voltage noise across the sample is amplified by two low-noise voltage preamplifier chains independently, each with the total gain 10,000 (LI-75 followed by SR-560), and recorded by a high-speed digitizer (NI-PCI5122) at a sampling rate of 5 MHz within 10 ms for each time series. The noise signal is very sensitive to the environment and a Faraday cage (dash line in Extended Data Figure 4**a**) is crucial to minimize interference from background electromagnetic signals. The voltage fluctuations in the two amplifier chains are cross-correlated to suppress contributions from amplifier noise (nominally uncorrelated between the two chains). The cross-correlation analysis finds the in-phase components between the two time series signals and gives the power spectral density of the correlated components. Each measurement of the power density spectrum of noise is an average of 4,000 of these cross-correlations, and it takes about 1.5 min.

Resistive and capacitive parasitic contributions are unavoidable in this measurement approach. While the parasitic series resistance (on the order of Ω) is negligible compared with the typical sample differential resistance (on the order of kΩ), the parasitic capacitance to ground (wiring plus the device itself) may affect the measured noise spectrum. For a standard treatment of capacitive attenuation of the voltage noise at high frequencies[38], the equivalent circuit diagram is shown as Extended Data Figure 4**b**. The voltage noise at the input end of the preamplifier is:

$$\tilde{v}^2 = v_{na}^2 + \frac{i_S^2 R_S^2}{1 + (R_S C_P \omega)^2}$$

where $i_S$ is the noise of the source current, $R_S$ is the (generally bias-dependent) differential resistance of the sample, $C_P$ is the parasitic capacitance to ground in the system, and $v_{na}$ is the input voltage noise of the preamplifier. After the amplification gain $G$, the voltage noise becomes $G(v_{na}^2 + \frac{i_S^2 R_S^2}{1+ (R_S C_P \omega)^2})$. An ideal cross correlation would eliminate the input voltage noise from the preamplifiers and the final expression for the measured power spectra density is:

$$\tilde{S}_V(\omega) = \frac{G S_V}{1 + (R_S C_P \omega)^2}$$

where $S_V = i_S^2 R_S^2$ is the intensity of the intrinsic voltage noise.

## 5. The $R_S C_P$ model and fitting the noise spectra

The $R_S C_P$ model gives an explicit expression for the spectral density of voltage noise power and can be used for fitting the measured spectrum. Extended Data Figure 5**a** shows an example spectrum of Johnson-Nyquist (J-N) noise from a fixed 2.17 kΩ resistor at $T$ = 300 K. Like the shot noise, J-N voltage noise is white noise, with power spectral density $S_V = 4k_B TR$, where $R$ is the ohmic resistance. The measured power spectral density decays with increasing frequency, and it is caused by the parasitic capacitance. It could be well fitted by the $R_S C_P$ model equation, see Extended Data Figure 5**a**, where the blue dots are the measured spectrum and the red dashed line is the fit. There are a few spikes in the spectrum, which come from unavoidable instrumental pickup of extrinsic environmental signals. A robust fitting algorithm[39] is used to minimize the contributions of those outliers and fit the underlying spectrum. With the same device at fixed temperature, the fitted parasitic capacitance is almost constant at different bias. The parasitic capacitance extracted from the fitting over all measurements of all devices is 300 ± 100 pF, which is reasonable considering all the input capacitance of the preamplifier and the parasitic capacitance in the transmission line. The variance mostly comes from the device-to-device variations in wiring or bonding.

The linear dependences of the spectral density of power of J-N noise on the resistance $R$ and temperature $T$ are verified in our experiments. Extended Data Figure 5**b** shows the spectral density of power of J-N voltage noise versus the resistance at 300 K. The linear relationship holds within a large resistance range, from 10 Ω up to 30 kΩ. Smaller resistors would make the J-N signal-to-noise ratio (SNR) too small to be accurately resolved, and larger resistances affect the voltage amplifier's gain and amplifier noise properties. This simple J-N linear dependence provides a calibration reference to our system. Similarly, the J-N linear dependence on temperature is also observed, from 300 K down to 5 K, see Extended Data Figure 5**c**. A small deviation is observed at lowest temperatures, where the junction's real temperature might be slightly different from the setting temperature of the cryo-station.

For a macroscopic diffusive conductor at constant temperature, the J-N noise should not depend on the applied dc voltage or current, which is also verified in our system. We applied current up to ±20 µA to a fixed resistor and found that the noise spectral densities are always consistent with theoretical expectations. This verifies that the current and voltage sources are clean and shows that any bias-dependence of the noise originates from the samples rather than from the measurement system.

The same basic approach is applied to the shot noise measurement in LSCO tunnel junctions. At each temperature, the dc current bias is finely swept in increments of about 700 nA up to ± 20 µA. The averaged cross-correlation spectrum is recorded at each bias and fit with the $R_S C_P$ model equation (the red dashed lines in Extended Data Figure 5), inserting for $R_S$ the measured differential resistance at a given bias. The spectral density of voltage noise power could be extracted from the fitting parameters. Together with the d$I$/d$V$ measurements at the same bias conditions, the voltage noise is translated to the current shot noise by $S_I = S_V/(dV/dI)^2$.

It is noticeable from the spectra that at high current or voltage bias, the 1/$f$ noise becomes more noticeable, as seen in the low frequency limit in Extended Data Figure 5**e-j**. Hence, in our data analysis we restrict the fitting range to frequencies sufficiently high to mitigate any effects of 1/$f$ contributions, as verified through consistency of the $R_S C_P$ model. The fit parameters for those data sets are shown in Extended Data Table 1.

This analysis procedure takes account for the non-Ohmic sample conduction in multiple ways. $R_S C_P$ fit at each bias is consistent with a stable wiring capacitance and the measured differential resistance at each bias. Each voltage noise power spectrum is converted into current noise power by using the differential resistance measured at each particular bias. This is analogous to the procedure employed by Cron *et al.*[40] used to examine shot noise and MAR in atomic-scale superconducting contacts. The comparison with the Poissonian noise expectation uses the measured (non-Ohmic) $I$ and $V$, as in Ref. 6.

## 6. Sample-to-sample variations

We measured two devices of each doping level from $x = 0.15$ to $x = 0.10$. In Extended Data Figure 6**a-h** the noise ratio is shown for the four devices featured in Fig. 1 of the main text, and in Extended Data Figure 6**i-p**, the results are shown for the other four devices. As in the main text, the shot noise intensity shows an enhancement above single-charge tunneling expectations above $T_c$ for each doping level. At temperatures far above $T_c$, the noise density is close to the prediction for single-electron tunneling, with the noise ratio close to 1.

We also observed some variance from sample to sample. For the $x = 0.15$ optimal doping sample, the noise ratio falls below 1 at high temperatures when the bias is larger than 10 mV. This might be related to the charge transfer issue as the doping level increases[32-35], which would indicate an increased barrier transparency and undermine the constant-barrier tunneling approximation. For the $x = 0.14$ doping sample in Extended Data Figure 6**b**, we observe atypically large enhancement of the noise ratio at temperatures below $T_c$. One possible explanation for enhanced noise response in this device relative to the others is a local variation in the barrier properties, as described above. These observations are strong motivations for future experiments to examine noise response with thinner LCO barriers, as well as to build on the work in ED section 2 and further study in depth the crystalline, chemical, and electronic structure of the barriers and interfaces using transmission electron microscopy and electron-energy loss spectroscopy with atomic-resolution.

## 7. Error analysis

The lock-in amplifier technique gives good accuracy in measuring the differential conductance, thus the uncertainty in the noise measurements mostly comes from systematic errors and the spectrum acquisition and fitting procedures.

Systematic uncertainty in the noise may originate from measurement calibration, temperature inaccuracy, digitizer reading errors, and cross-correlation residuals. For the calibration process, the linear fitting in Extended Data Figure 5**b** has $R^2 = 0.9998$ and the uncertainty is less than 1%. The temperature is controlled by the PPMS PID feedback system. Normally the device's temperature is stabilized in a few minutes and the temperature accuracy is within 20 mK. Because of the need to isolate the noise measurement electronics from the PPMS ground, with our home-built shot noise probe, at lowest temperatures (below 10 K), the PPMS cooling power transmitted to the sample is limited; zero-bias J-N noise for nominal cryostat temperature of 5 K indicates a sample temperature of 6 K. These issues are negligible at temperatures above 10 K. The PCI-5122 digitizer's accuracy is within 0.65% in our input range, which is negligible compared to other error sources. When the signal to noise ratio (SNR) is large, the cross-correlation could restore the original signal with very good accuracy. In the situation of extremely low temperature/small resistance, the SNR is low and the cross-correlation accuracy is affected. In our experiment, at the lowest cryostat temperature for the specialized noise probe (5 K), the Johnson-Nyquist noise at zero bias accuracy is within 15% of expectations based on the measured differential resistance. As temperature increases to 20 K and above, the measured noise at zero bias is consistent with the Johnson-Nyquist noise to better than 3%, indicating very good temperature accuracy. Overall, the typical standard deviation for the noise ratio in LSCO devices due to systematic errors is 0.015.

Converting the noise ratio as a function of temperature and bias into the inferred paired contribution fraction is done assuming $S_I = (1-z)2eI \coth(eV/2k_BT) + z\, 4eI \coth(eV/k_BT)$. This assumes that at any given bias and temperature there is a noise contribution due to single-charge tunneling of the form $2eI \coth(eV/2k_BT)$ and a contribution due to pairs such that $q^*(V) = 2e$, given by $4eI \coth(eV/k_BT)$. This is consistent with prior analyses used for Andreev reflection (e.g., Ref. 6). Note that $S_I$ reverts to the Johnson-Nyquist expectation $4k_BT(I/V)$ as $V \rightarrow 0$, regardless of the pair fraction $z$. This is a consequence of the fluctuation-dissipation theorem, and this implies that in the zero-bias limit of equilibrium, it is not physically possible to extract $z$ from the noise.

Rearranging gives $z = [(S_I/S_{I,e}) - 1]/[2 \coth(eV/k_BT)/\coth(eV/2k_BT) - 1]$, where $(S_I/S_{I,e})$ is the noise ratio. When $eV = k_BT$, the denominator of that expression is approximately 0.2135. When $eV = 0.3\, k_BT$, the denominator is approximately 0.02217. (Consistent with the fluctuation-dissipation expectations, small experimental uncertainties in the noise ratio become infinite uncertainties in $z$ as $V \rightarrow 0$.) The typical systematic uncertainty in noise ratio of +/- 0.015 translates into an uncertainty in $z$ of +/- 0.070 when $eV = k_BT$, and +/- 0.68 when $eV = 0.3\, k_BT$.

We note that there is a difference between the fraction of current contributed by paired carriers and the fraction of all carriers that are paired. The measurement is of those carriers that tunnel along the *c*-axis, though the current is dominated by carriers from the antinode portion of the Fermi surface. [28,41]

For the spectrum acquisition process, longer averages would help narrow the spectrum distribution, see Extended Data Figure 7. To analyze this dependence, a relatively flat spectrum region is selected and then normalized with its mean value (Extended Data Figure 7**a**). The distribution of the normalized power spectral density (PSD) with different averaging time is plotted and fitted with a Gaussian distribution curve, see Extended Data Figure 7**b**,**c**. The standard deviation of this distribution represents the variance of the spectrum density. In our experiment, we used 96 s average (~4,000 times), and the standard deviation is about 2.5%.

Similar with Extended Data Figure 5**d**, with a fixed average time of 96 s, we also took the Johnson-Nyquist noise spectrum repeatedly to estimate the spectrum collecting and fitting error. The standard deviation of 50 fitted PSDs is 2.3%.

## 8. Conventional SIS junction

We have performed analogous noise measurements on a Nb/AlOx/Nb tunnel junction, available commercially from STAR Cryoelectronics. The junction is fabricated on a doped Si substrate, and the critical temperature of the Nb electrodes is approximately 9 K. The differential conductance and noise of the device are shown in Extended Data Figure 8**a-d**, while Extended Data Figure 8**e** shows the results when $z$ is extracted from the data, following the same procedures as for the cuprate devices. Because of the comparatively low junction resistance, the bias range is restricted by limitations on the measurement current, and contributions of 1/f noise that grow quadratically with bias current. The low junction resistance also corresponds to a higher amplifier noise contour for the first-stage LI-75 amplifiers in Extended Data Figure 4**a**, compared with the higher resistance LSCO devices. Noise measurements within the gap bias range in the superconducting regime in this structure are obscured by the presence of Josephson current in the device below $T_c$ and resulting enhanced environmental pickup.[42] Similar enhanced environmental pickup is seen at the lowest temperatures in the LSCO devices.

## 9. Noise as a function of bias current

Shot noise measurements in the literature are often plotted as a function of bias current rather than voltage, because such measurements are frequently in the high bias regime ($eV \gg k_B T$) where the expected Poissonian current noise takes on the simple limiting form $S_I = 2eI$. Data in the main text are presented as a function of bias voltage to facilitate comparison with the gap energy scale of superconductivity, but plotting the noise as a function of bias current also shows the essential features (enhancement above the expected Poissonian value at high biases and elevated temperatures). Extended Data Figure 9**a-e** shows the data for Fig. 3 replotted as a function of bias current.

## 10. Multiple Andreev Reflection (MAR) and enhanced noise at low bias in the superconducting regime

The large noise enhancements observed at low bias and below $T_c$ are reminiscent of multicharge tunneling via higher order Andreev reflection processes. MAR has been reported in SIS structures,[6,7,43,44] and while coherence is not required for Andreev processes, barrier transparency plays a critical role in the magnitude of the effect. While lacking a detailed theoretical prediction for this particular situation (*d*-wave order parameter, *c*-axis tunneling with preservation of transverse momentum), it is possible to compare the enhanced noise peaks with a simple model.

As different multiple Andreev charge transfer processes are kinetically allowed depending on the bias, the expected effective charge is bias-dependent ($q^* = ne$ for $2\Delta/n < eV < 2\Delta/(n-1)$ for $n = 2, 3, ...$). Extended Data Figure 9**e,f** shows a finite temperature expectation for the noise and noise ratio as a function of bias, $V$, using $S_I = 2q^*(V)I \coth(q^*(V)V/2k_B T)$ with this assumption for $q^*$ as the comparator to Poissonian single-charge tunneling, along with the data at 5 K for the sample used in Figure 3. The observed enhanced noise peaks differ in detail from the simplified MAR expectations.

**Methods References**

**Data Availability**

The data used to produce the figures in the main text as well as in the Extended Data are provided with the paper. Data are also available online via a link attached to this article through the SpringerNature Research Data Support Service.

# Extended Data Figure Legends

**Extended Data Figure 1**: **Device fabrication process**. **a**. LSCO/LCO/LSCO film is grown on top of LSAO substrate with a thin layer of in situ deposited Au covering the film. **b**. The film is etched into about 20 μm sized bars defined photolithographically. This is a deep etch all the way into the substrate. **c**. A second dry etch step removes part of the top LSCO and middle LCO layers, and stops in the middle of the bottom LSCO layer, creating 10-20 μm sized mesas. **d**. A thick layer of $Al_2O_3$ (100 nm) is evaporated to isolate the future top Au contact (150 nm) and bottom Au contacts, to avoid parallel conduction paths. **e**. Contacts are defined lithographically and Au is evaporated to make contact with top and bottom LSCO layers. **f**. A false-colored SEM image of the device. **g**. STEM cross-section of a representative device structure, showing the atomic perfection of the ALL-MBE process.

**Extended Data Figure 2**: **Transport in LSCO-LCO-LSCO ($x = 0.15$) film and tunnel junction properties**. **a**. $R$-$T$ measurement on the Hall-bar device fabricated in this film shows the superconducting transition temperature $T_c = 38$ K. **b**. Tunneling differential conductance in a trilayer junction fabricated in this film. **c**. Log-log plot of the I-V characteristics of two $x = 0.15$ tunnel junction devices, demonstrating device-to-device reproducibility and lack of any supercurrent down to pA levels at dilution refrigerator temperatures.

**Extended Data Figure 3**: **Bias-dependent noise and differential conductance**. Noise data from Figure 3 reproduced with accompanying un-normalized differential conductance data.

**Extended Data Figure 4**: **Electrical circuit diagrams for the shot-noise measurement setup**. **a**. The diagram of the two channel cross-correlation method. **b**. The equivalent circuit diagram can be modeled as an $R_S C_P$ circuit, where $i_S$ is the noise source, $R_S$ is the (bias-dependent) differential resistance of the sample, $C_P$ is the parasitic capacitance in the system, and $v_{na}$ is the input voltage noise of the preamplifier.

**Extended Data Figure 5**: **$R_S C_P$ model fitting, noise power spectra density calibration and example spectra of an LSCO tunnel junction**. **a**. The spectrum of the power density of Johnson-Nyquist (J-N) voltage noise in a 2.17 kΩ resistor at $T = 300$ K, measured by the cross-correlation method. The red line is fitting based on the $R_s C$ model. **b**. J-N voltage noise of various resistors at 300 K. The voltage noise $S_V$ has a simple linear dependence on the resistance of the resistor that is used as a calibration reference. **c**. The J-N noise is also linearly dependent on temperature for a fixed resistor (2.17 kΩ). **d**. For a fixed resistor (2.17 kΩ), the J-N noise is independent of the bias current, as expected for a macroscopic diffusive conductor. **e-h**. Example spectra of an LSCO

tunnel junction for $x = 0.15$, recorded at $T = 50$ K. The dc bias current is marked for each panel. Red dash line are fits based on the RC circuit model. The sharp spikes result from environmental pickup of specific frequencies, and the fitting procedure is not influenced by these. Such environmental pickup is larger at the lowest temperatures below $T_c$.

**Extended Data Table I**: **Example fitting parameters**. The $R_SC_P$ model fitting parameters for the noise spectrum in Extended Data Figure 5e-j.

**Extended Data Figure 6**: **Noise ratios as a function of temperature and bias**. The noise ratio for the 4 LSCO devices featured in Fig. 1 at various doping levels as indicated, measured below $T_c$ (**a-d**) and above $T_c$ (**e-h**). The noise ratio for the other 4 LSCO devices at various doping levels as indicated, measured below $T_c$ (**i-l**) and above $T_c$ (**m-p**).

**Extended Data Figure 7**: **The variance in power spectral density (PSD) with different averaging times**. **a**. A relatively flat region (red) is selected to analyze the distribution of variations in the PSD. Sharp spikes are environmental pickup of discrete frequencies; these are not used in the fitting procedure. **b, c**. The normalized PSD distribution in the selected region for a 96 s average and a 6 s average. The red line is the Gaussian fit to the distribution. **d**. The standard deviation of the distribution for different averaging times.

**Extended Data Figure 8**: **Shot noise in a Nb tunnel junction**. **a-d.** Noise measurements (blue points with error bars) and differential conductance (green) as a function of bias and temperature for a commercial Nb/AlO$_x$/Nb tunnel junction that exhibits Josephson supercurrent below $T_c = 9$ K. **e**. Inferred pair fraction $z$ as a function of bias and temperature for this device. Red dash-dot line: the superconducting gap region outside which one would expect $z = 0$ from the BCS theory for the measured value of $T_c$. Green dashed line: $V = k_BT/e$. As $eV/k_BT \to 0$, discrimination of $z$ via noise measurements is not possible (see Methods). Grey region indicates where uncertainty in $z$ exceeds 0.5.

**Extended Data Figure 9**: **Noise as a function of current, and comparison with Andreev reflection**. Noise data from Fig. 3 plotted as a function of bias current rather than bias voltage and the noise data at 5 K compared with expectations of a very simplified model of multiple Andreev reflection. **a-d.** The dashed red line shows the single-charge tunneling Poissonian expectation $2eI \coth(eV(I)/2k_BT)$, based on the measured $I(V)$ at each temperature. **e-f.** The red traces assume a bias-dependent effective charge based on kinetically allowed Andreev processes ($q^* = ne$ for $2\Delta/n < eV < 2\Delta/(n-1)$ for $n = 2, 3, \ldots$) for a fixed isotropic gap, $\Delta$, combined with a finite temperature expectation for the noise $S_I = 2q^*(V)I \coth(q^*(V)V/2k_BT)$.

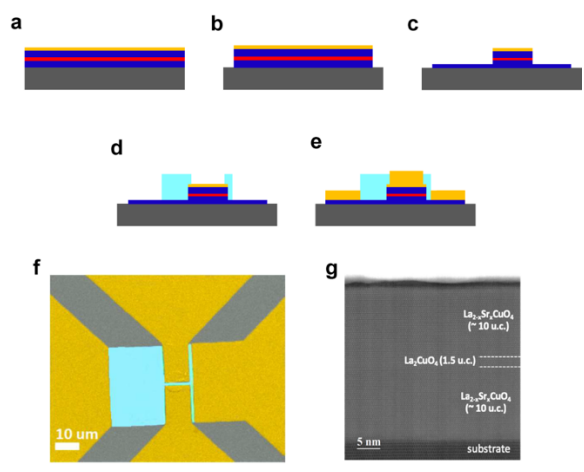

Extended Data Figure 1

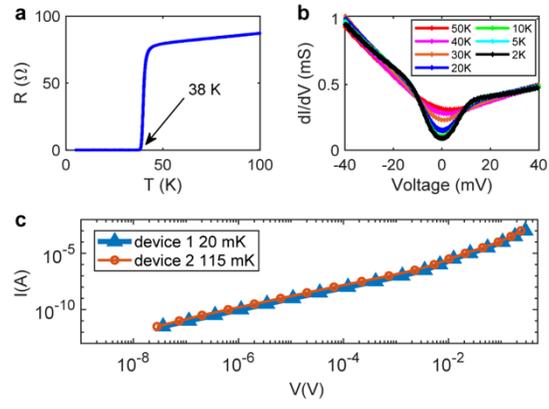

Extended Data Figure 2

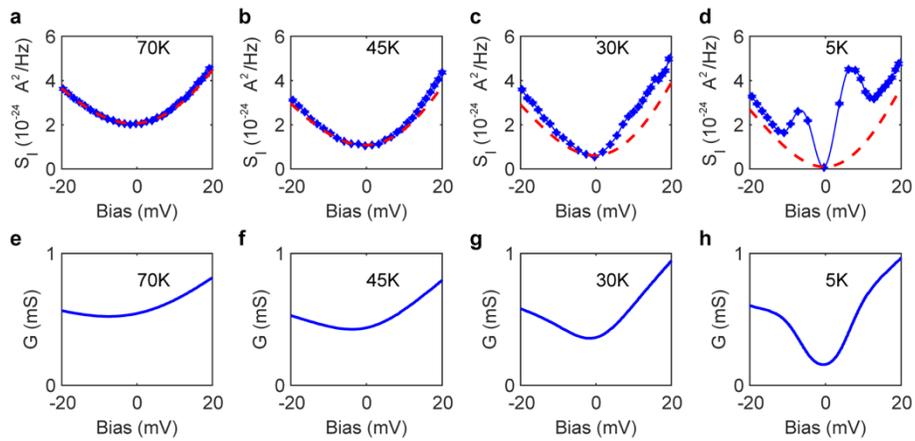

Extended Data Figure 3

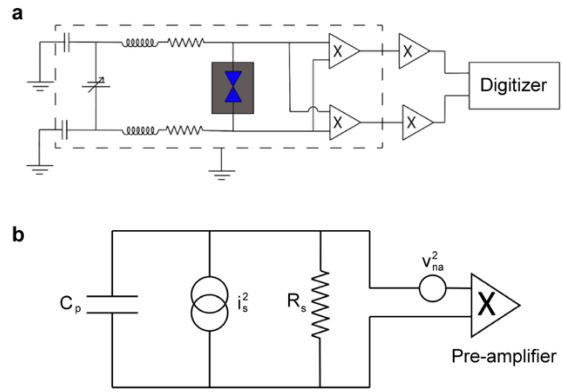

Extended Data Figure 4

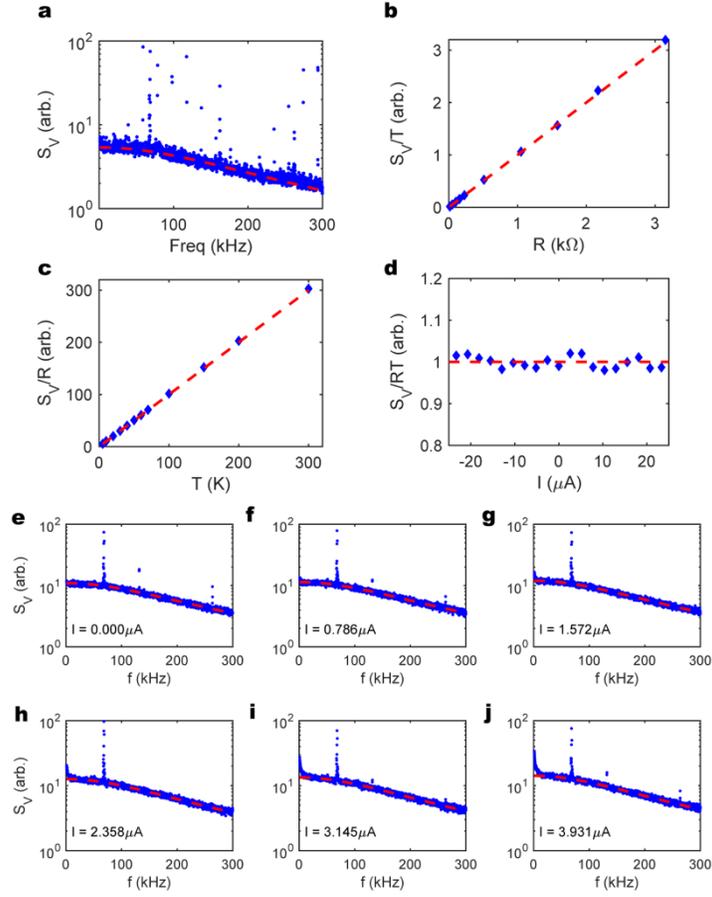

Extended Data Figure 5

| Bias Current (μA) | $S_V$ (arb.) | $2p\, R_S C_P$ (1/Hz) | $dI/dV$ (S) |
|---|---|---|---|
| 0 | $9.55 \times 10^{-8}$ | $5.16 \times 10^{-6}$ | $4.17 \times 10^{-4}$ |
| 0.786 | $9.87 \times 10^{-8}$ | $4.93 \times 10^{-6}$ | $4.10 \times 10^{-4}$ |
| 1.572 | $1.06 \times 10^{-7}$ | $5.22 \times 10^{-6}$ | $4.06 \times 10^{-4}$ |
| 2.358 | $1.13 \times 10^{-7}$ | $5.07 \times 10^{-6}$ | $4.07 \times 10^{-4}$ |
| 3.145 | $1.22 \times 10^{-7}$ | $5.38 \times 10^{-6}$ | $4.12 \times 10^{-4}$ |
| 3.931 | $1.32 \times 10^{-7}$ | $5.08 \times 10^{-6}$ | $4.15 \times 10^{-4}$ |

Extended Data Table 1

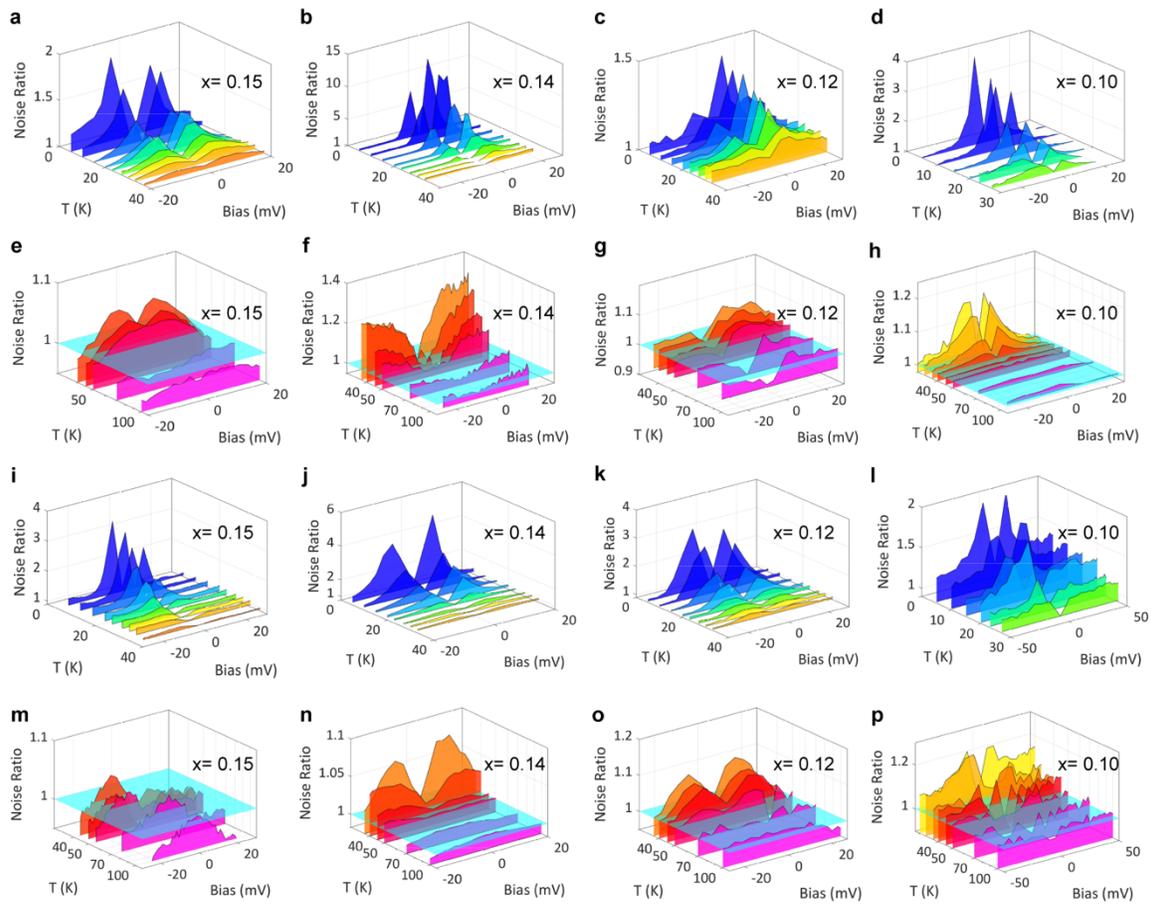

Extended Data Figure 6

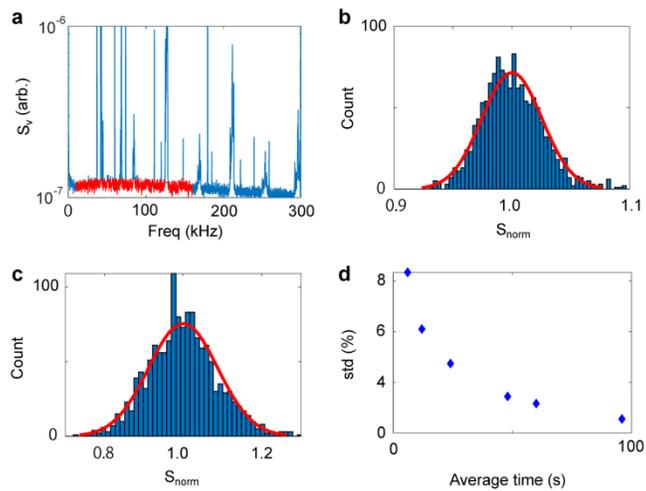

Extended Data Figure 7

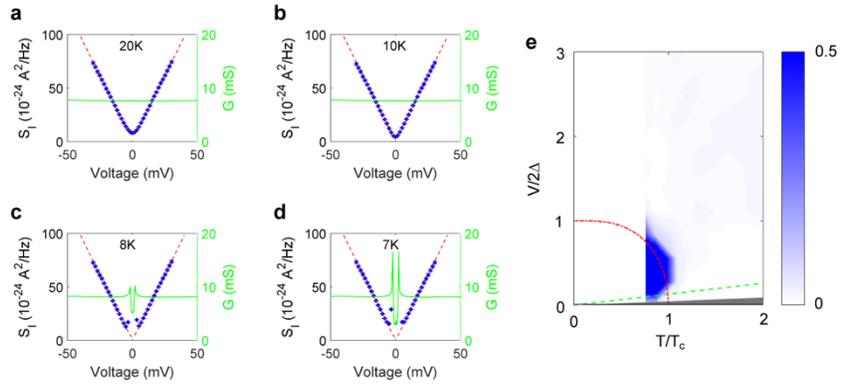

Extended Data Figure 8

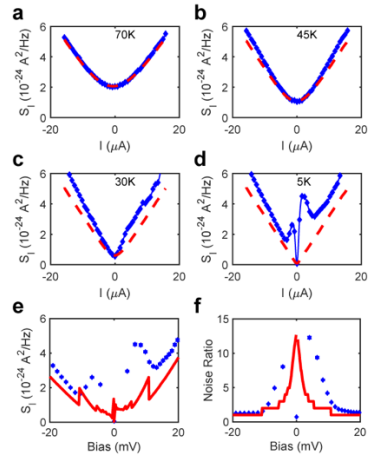

Extended Data Figure 9